# Reduced Muscle Fatigue Using Continuous Subthreshold Kilohertz Stimulation of Peripheral Nerves

Long Meng, Paola Terolli, and Xiaogang Hu, *Senior Member, IEEE*

*Abstract*—Functional electrical stimulation (FES) is a prevalent technique commonly used to activate muscles in individuals with neurological disorders. Traditional FES strategies predominantly utilize low-frequency (LF) stimulation, which evokes synchronous action potentials, leading to rapid muscle fatigue. To address these limitations, we introduced a subthreshold high-frequency (HF) stimulation method that employed continuous, charge-balanced subthreshold current pulses at kilohertz frequencies, designed to evoke motor unit (MU) activation similar to voluntary activation. We evaluated the effectiveness of HF stimulation on the reduction of muscle fatigue across different force levels (10 %, 25 %, and 40 % of maximum force). The HF stimulation utilized continuous charge-balanced, short pulses of 80 μs (at a 10 kHz frequency) targeted the ulnar/median nerve bundles. We compared the fatigue effects with conventional LF stimulation and voluntary muscle contractions. Our results indicated that HF stimulation maintained more sustained force outputs and muscle activation over a prolonged time compared with LF stimulation. The HF stimulation also evoked a more dispersed muscle activation pattern, similar to voluntary muscle contractions. These findings suggest that HF stimulation can significantly enhance the sustainability of muscle contractions and reduce muscle fatigue, potentially improving the efficacy and applicability of FES in clinical and home-based settings for individuals with neurological impairments.

## I. Introduction

FUNCTIONAL electrical stimulation (FES), electrical stimulation of the nervous system for functional purposes, is widely used in both clinical and home settings to enhance muscle strength and assist motor functions in individuals with neurological disorders [1], [2], [3], [4]. Typically, stimulation electrodes are placed around major peripheral nerve bundles to drive a range of muscles [5], or directly on specific muscle bellies near the innervation zone that activate the distal ends of motor axons [6], in order to facilitate hand motions [7]. However, the clinical adoption of conventional FES faces several challenges. One major issue is the rapid onset of muscle fatigue evoked by FES. Conventional FES generally activates the target muscles in non-physiological manners. Although activations in random orders have been observed, which may be attributed to variations in the distances between the electrodes and branches of motor nerves [8], the order of recruitment through electrical stimulation generally differs from the natural physiological order of recruitment [8], [9]. Furthermore, axons with different activation thresholds may be evoked by a single stimulus pulse, leading to a highly synchronous activation of axons [8]. Consequently, the evoked forces are time-locked to the timing of the stimulation pulses. In contrast, the delay between axon activations is relatively prolonged with asynchronous firing of motor units (MUs) during voluntary activations. With the conventional stimulation strategy, the non-physiological axonal activation (such as non-physiological recruitment order and temporal synchrony of activation that is time-locked to the stimulation pulses) may quickly cause muscle fatigue onset [10], [11], [12] and a large force decline [13], potentially hindering the wide application of the FES systems.

To address these issues, various approaches have been investigated in previous studies. For instance, Buckmire et al. demonstrated that distributed stimulation with interleaved activation across electrodes could achieve sustained force output [14]. Their findings highlight the potential benefits of spatially-patterned stimulation to mitigate fatigue while maintaining functional force levels. Similarly, another study by Buckmire et al. showed that nerve stimulation with feedback-controlled stimulation intensity could sustain force output comparable with voluntary contractions during prolonged activation [15]. Moreover, several studies employed spatially-patterned stimulation approaches through multi-channel stimulation to activate different muscle areas by targeting specific electrode groups [15], [16], [17]. Electrodes were switched during stimulation to achieve the asynchronous

This study was supported in part by the National Science Foundation (CBET-2246162, IIS-2319139). *(Corresponding author: Xiaogang Hu.)*

Long Meng and Paola Terolli (deceased) are with the Department of Mechanical Engineering, Pennsylvania State University-University Park, PA, USA (e-mail: lmm7405@psu.edu; paola.terolli@psu.edu).

Xiaogang Hu is with the Department of Mechanical Engineering, Pennsylvania State University-University Park, PA, USA, and also with the Departments of Kinesiology, and Physical Medicine & Rehabilitation, the Huck Institutes of the Life Sciences, and the Center for Neural Engineering, Pennsylvania State University-University Park, PA, USA (e-mail: xxh120@psu.edu).

activation of different muscle areas, thereby delaying muscle fatigue. Similarly, some studies [18], [19], [20] placed invasive or non-invasive electrode arrays near nerve bundles. Although these techniques can reduce muscle fatigue, a group of nerve fibers may still be synchronously activated by a specific channel in the multi-channel array, producing compound action potentials that are typically characterized by large amplitudes and distinct phases [19], [21]

The axonal recruitment pattern can also be affected by the temporally-patterned stimulation. Compared with the conventional stimulation in low frequency (LF), the high-frequency (HF) stimulation with bursts at kHz frequencies can evoke transient changes in axonal firing patterns, such as transient conduction block [22] or asynchronous firings [23]. By modulating the burst amplitude of stimulation or continuously delivering stimulation pulses at kHz frequencies, axons exhibit random asynchronous firing patterns, caused by different axonal response characteristics [24]. To effectively evoke muscle activity with asynchronous activation, previous studies [13], [25], [26], [27] have explored patterned stimulation with short electrical pulses delivered in bursts at kHz frequency, and each pulse only caused subthreshold depolarizations of the axonal membrane potentials. The pulses can also be amplitude-modulated through sinewave patterns or through ramps to recruit fibers. Axons with different diameters would require varying numbers of pulses to reach their activation thresholds, resulting in asynchronous axon activations. This stimulation strategy could also activate substantial afferent fibers, thereby leading to orderly MU recruitment through the reflex pathways. The asynchronous axon activations through suprathreshold kilohertz spinal stimulation have been proven to mediate pain relief by blocking the transmission of spikes to the somatosensory cortex through feedforward inhibition [28]. In a recent study, when bursts of current pulses were delivered continuously, the evoked muscle activation pattern was similar to voluntary contraction [7]. However, it is unclear regarding the effectiveness of this stimulation approach in reducing muscle fatigue across a range of force levels for practical use.

In this study, we evaluated the efficacy of a recently developed, temporally-patterned stimulation approach in reducing muscle fatigue over a wide range of force levels (10 %, 25 %, and 40 % of maximum force), which are sufficient to satisfy most daily tasks. Specifically, we applied continuous, charge-balanced stimulation (80% duty cycle) with bursts of short pulses (pulse width of 80 μs and pulse interval of 20 μs) at a 10 kHz carrier frequency (termed HF stimulation) to the ulnar/median nerves. We also evaluated the fatigue effect of the conventional stimulation with a pulse width of 500 μs at a low frequency of 30 Hz (termed LF stimulation) for comparison. Additionally, we evaluated the fatigue effect under the voluntary contraction (termed Vol) for each condition, serving as a baseline for evaluating the efficacy of our stimulation approach. Our findings revealed that the forces generated by our developed HF stimulation declined more slowly, and the sustained force level was higher compared with the low-frequency stimulation. The muscle activation strength decreased over time under LF stimulation. In contrast, the muscle activation variation under the HF stimulation tended to be stable, closer to that under the voluntary contraction. Additionally, similar to the voluntary contraction, we can achieve a more dispersed spatial distribution of muscle activation using the HF stimulation than that using the LF stimulation. These results reveal that we can effectively reduce muscle fatigue, which can advance FES techniques with a potential to evoke prolonged muscle contraction, making them feasible for daily tasks that require sustained motor output at different force levels.

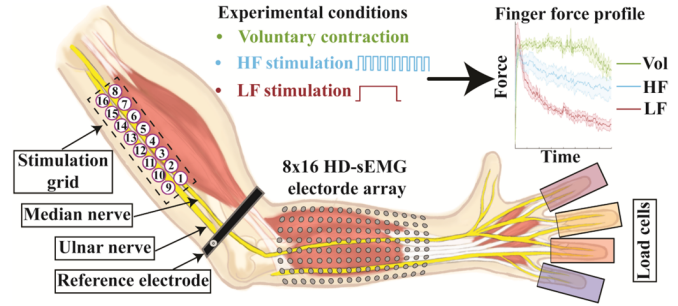

Fig. 1. Experimental setup. Specific muscles were activated by a stimulation grid of electrodes placed along the median and ulnar nerves under three experimental conditions. The evoked motor activity was recorded using a HD-sEMG grid and four individual load cells.

## II. METHODS

### A. Subject Information

Eight healthy individuals (four males and four females, aged between 25 and 38 years) without any history of neurological disorders participated in the study. All participants provided informed consent, which was approved by the Institutional Review Board of the Pennsylvania State University (Approval Number: STUDY00021035).

### B. Experimental Setup

Participants were comfortably seated in a height-adjustable chair during the experiment. Their hands were secured with foam pads on both the dorsal and palmar sides, and a soft foam pad supported their forearms. As shown in Fig. 1, four fingers were comfortably attached to miniature load cells (SM-200N, Interface, Scottsdale, AZ) using Velcro straps for accurate force measurement. These load cells independently recorded finger flexion forces at a sampling rate of 1 kHz.

We used a high-density surface electromyogram (HD-sEMG) electrode array to capture activation patterns of the extrinsic finger flexor muscles. Before the sEMG electrode placement, we cleaned the forearm skin using 70% isopropyl alcohol pads to reduce the skin-electrode interface impedance. Then, an 8 × 16 channel sEMG electrode array (inter-electrode distance: 10 mm, electrode diameter: 3 mm) was attached to the finger flexor muscles (Fig. 1). We determined the placement of the electrode array by palpating the forearm flexor muscles as subjects flexed their fingers voluntarily. The center of the electrode array was then aligned with the midpoint between the styloid processes and the olecranon. To minimize stimulation artifacts, we placed a reference electrode on the wrist, and a common ground electrode on the elbow. Monopolar sEMG signals were recorded using the EMG-USB2+ system (OT Bioelettronica, Torino, Italy) at a sampling rate of 2048 Hz, a gain of 200, and with a band-pass filter ranging from 10 to 900 Hz. The collected sEMG signals were preprocessed by an sEMG-specific motion artifact removal approach [29].

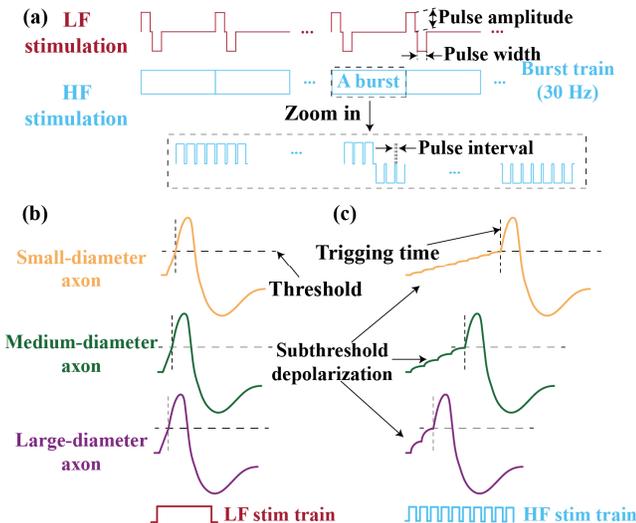

Fig. 2. Stimulation protocols. (a) LF and HF stimulation waveforms. The waveform of the conventional LF stimulation was charge-balanced, delivered at 30 Hz. Specifically, the LF stimulation started with positive rectangular waves, followed by negative rectangular waves of the same pulse amplitude and duration. The HF stimulation was delivered in a form of bursts at 30 Hz. Each burst contained short charge-balanced pulses, including a series of positive short pulses followed by an equal number of negative pulses with the same absolute amplitude. The dashed rectangle highlighted a zoom-in view of the patterned pulses within a single burst in the HF stimulation. The hypothesized fiber responses using the LF and HF stimulation are presented in (b) and (c), respectively. In (b), the action potentials of different axons were evoked by an individual LF pulse in a highly synchronous manner despite the variability in their individual axial resistances. In (c), an action potential can be triggered by the accumulation of subthreshold depolarizing responses evoked by the short HF pulses. The numbers of subthreshold depolarization differ for the axons with different diameters to trigger action potentials, thereby resulting in asynchronous activations. The vertical dashed lines overlaid on the action potentials in (c) and (d) indicate the triggering time of an action potential, which is the moment when the membrane potential reaches the threshold, and an action potential is generated.

1) **Stimulation Setup**

We delivered the electrical stimuli using sixteen gel-based electrodes, each with a diameters of 1 cm, arranged in a 2x8 grid. As shown in Fig. 1, these electrodes were placed on the medial side of the upper arm directly beneath the short head of the biceps brachii, targeting the area where the ulnar and median nerves run close to the skin surface. This placement was aligned with a vector connecting the center of the axilla to the medial epicondyle of the humerus, ensuring the electrodes were well-positioned along these nerve paths beneath the skin surface. Earlier studies have demonstrated that various hand grasp patterns can be achieved by using bipolar stimulation to activate different groups of axons [5], [30]. Then, these electrodes were connected to the columns of a switch matrix (Agilent Technologies, Santa Clara, CA), the rows of the switch were linked to the anode and cathode of a multi-channel stimulator (STG4008, Multichannel Systems, Reutlingen, Germany). Correspondingly, we developed a MATLAB user interface to control stimulation trains with varying parameters and to enable the delivery of electrical stimuli to any selected pairs of electrodes, facilitating the identification of electrode pairs capable of inducing the targeted muscle contractions. We evaluated the index force of one subject, middle finger force of three subjects, and the index-middle finger force of four subjects. In the experiment, we evaluated muscle fatigue over time under three initial flexion force levels (10 % maximum voluntary contraction (MVC), 25 % MVC, and 40 % MVC) with voluntary contraction, LF stimulation, and HF stimulation conditions, respectively. We selected 40 % MVC as the upper limit because it can satisfy the demands of most daily activities.

In the LF stimulation experiment, a single biphasic rectangular pulse (Fig. 2(a)) was delivered to the selected pair of electrodes per period (1/frequency). The stimulation frequency was set at 30 Hz, and the pulse width was 500 μs [7]. Before each trial, the stimulation current amplitude was adjusted to evaluate the finger force (10 % MVC, 25 % MVC, or 40 % MVC) evoked by the stimulation. If the evoked force matched the target finger force as designed, the trial proceeded with the adjusted current amplitude. The stimulation current amplitudes at different conditions are detailed in Table I. It is expected that the evoked action potentials of different axons are time-locked under the LF stimulation (Fig. 2(b)).

In the HF experiment, we delivered charge-balanced bursts of stimulation at a base frequency of 30 Hz for the HF stimulation. As shown in Fig. 2(a), the first half of each burst included positive rectangular waves with a fixed pulse width of 80 μs and pulse interval of 20 μs, and the second half shared the same parameters but with negative amplitudes. As a result, the individual pulses were continuously distributed across the entire stimulation period, maintaining an 80% duty cycle. Correspondingly, this parameter setting resulted in a pulse frequency of 10 kHz. We selected 10 kHz as the stimulation frequency based on previous studies [7], [27], which demonstrated there were no significant differences in evoking asynchronous activation of nerve fibers in the previously investigated kHz frequency range. Similar to the LF stimulation, we adjusted the stimulation amplitude (Table I) to reach the target force level at the beginning. Then, the stimulation amplitude was fixed for this force level to evaluate the effect of muscle fatigue reduction using the HF stimulation. A two-way repeated-measures ANOVA (factors: stimulation frequency × force level) revealed no significant interaction effect on the stimulation amplitudes ($F(2,14) = 1.044, p = 0.13$). However, the stimulation amplitude in the LF condition was significantly larger than that in the HF condition ($F(1,7) = 40.471, p < 0.001$). Our preliminary tests indicated that no recordable action potentials were triggered using individual narrow pulses unless those pulses were delivered in bursts at kHz frequency, suggesting that only subthreshold depolarization of the axonal membrane potential was evoked with individual pulses. Accordingly, the axon recruitment time is expected to be more dispersed compared with that using the LF stimulation. Namely, the delay between axon recruitments is increased using the HF stimulation (Fig. 2c). This is because axons with different diameters required different numbers of summations of subthreshold depolarizations to be activated [7], [27], as larger diameter axons exhibit greater subthreshold depolarizations due to their lower effective axial resistance, allowing them to reach the activation threshold earlier compared with small-diameter axons [31].

TABLE I
STIMULATION AMPLITUDE (MEAN ± STANDARD ERROR) ACROSS SUBJECTS

| Force Level | 10 % MVC | 25 % MVC | 40 % MVC |
|---|---|---|---|
| LF Amplitude (mA) | 5.22±0.68 | 5.76±0.81 | 6.13±0.74 |
| HF Amplitude (mA) | 3.55±0.74 | 4.01±0.79 | 4.04±0.61 |

2) **Experiment Procedure**

We first derived the flexion MVC of each finger by providing subjects with a 10-s trial to perform multiple maximum effort attempts. During each attempt, subjects were required to maintain their maximum effort for at least 2 s. The average values of these stable plateaus were calculated, and the highest average value was recorded as the MVC. This approach ensured a reliable estimation of the MVC. Then, we conducted a grid search to find a cathode-anode pair of stimulation electrodes that evoked flexion of at least one specific finger with minimal wrist activation and without significant or intolerable pain for the LF and HF stimulation conditions. The pulse amplitude was then adjusted to evoke the force levels. Specifically, three force levels (i.e., 10 % MVC, 25 % MVC, and 40 % MVC) were evaluated for three force generation strategies (i.e., voluntary contraction, LF stimulation, HF stimulation). Thus, a total of 9 trials (3 force levels × 3 activation strategies) were conducted for each subject, and the order of the trials were randomized. During the stimulation trials, subjects were required to avoid any voluntary contractions to ensure accurate measurement of evoked forces. During the voluntary trials, subjects were instructed to maintain the prescribed forces as long as they can, because they had difficulty maintaining a fixed voluntary effort during sustained contraction. Moreover, subjects were provided with visual feedback during the submaximal voluntary contractions to help them maintain the target force levels. Note that the voluntary trial would lead to a slower force decay than the stimulation trials with a fixed stimulus intensity, because the voluntary drive or effort increased over time to compensate for the fatigue effect. Between trials, we provided sufficient rest time until subjects indicated that they had recovered from the muscle fatigue, with an average rest time of 712 ± 59 s. Early work [32] has shown that the recovery time course of fatigue varies widely and depends on multiple factors, including the intensity and duration of the fatiguing exercise and MU composition of the targeted muscles. Therefore, to verify that the muscle fatigue is mostly recovered, subjects performed their MVC before trials, and the new trial continued if the measured forces matched the initial MVC. Considering the muscle tends to be fatigued faster under a higher level of force, we assigned a trial duration of 5 min, 4 min, and 3 min for trials of 10 % MVC, 25 % MVC, and 40 % MVC force levels, respectively. The shorter trial duration in higher force levels also reduced discomfort for participants.

*C. Data Analysis*

1) **Finger Force Analysis**

To reduce high-frequency noise of the recorded finger force signals, a moving-average filter was applied. The filter operated with a fixed window length of 1 s and a step of 0.5 s for subsequent computations. Additionally, we compared the averaged forces over different time periods. Considering that most of the finger forces were in the rising phase during the initial 5 s, the forces of initial 5 s were removed from the analysis. Due to the large variation in finger force during the early stage, we averaged the finger forces of smaller time intervals.

2) **sEMG Signal Analysis**

To quantify the sEMG signals, we first employed a stimulation-artifact removal approach [7] to remove stimulation artifacts from sEMG signals.

As shown in Fig. 3, LF stimulation artifacts were identified based on the stimulation timing. Specifically, the initial stimulation artifact was located as the peak amplitude within the first interstimulus interval. Subsequent artifacts were identified as the peak amplitudes within time windows centered on the stimulation times (previous stimulus time plus one interstimulus interval of 1/30 s), with a ±2.5 ms margin to account for potential artifact timing variations. Then, we centered a 5ms window on each stimulation artifact and replaced the data in this interval with random baseline sEMG signals (data collected in rest state without identifiable action potentials or stimulation). For the HF stimulation artifact removal, we segmented the sEMG signals using a fixed-size sliding window technique (sliding window size: 0.5 s, sliding step size: 0.5 s) (Fig. 4(e)). Considering the relatively large amplitude or time fluctuation of the artifact over 4 s [7], we processed the stimulation artifact in individual 4 s of signals (Fig. 4 (a), (b) and (c) for the HF). We then grouped every 8 consecutive sliding windows sequentially. For each 4-s sEMG signal, we extracted the shape of the stimulation artifact across the eight 0.5-s sEMG signals using the spike triggered averaging technique [33]. Specifically, we first aligned the sEMG of each channel across the eight 0.5-s sEMG signals. Then, we extracted the stimulation shape by computing the average response across the eight 0.5-s sEMG signals for each sEMG channel. During each 0.5-s HF stimulation train, the artifact was aligned in both time and magnitude for comparison with the original sEMG signal, which were adjusted by minimizing the mean squared error between the original sEMG signal (including artifacts) and the estimated artifact waveform (Fig. 4(f)). The adjustment was crucial because the amplitude of the stimulation artifacts ($A_{artifact}$) was several orders of magnitude greater than that of the sEMG signals ($A_{sEMG}$), i.e., $A_{artifact} \approx 10^n \cdot A_{sEMG}$. Then, as shown in Fig. 4(c) and (g), we minimized the artifact by subtracting it from the corresponding sEMG channel of eight 0.5-s sEMG signals, and conducted an outlier smoothing procedure to automatically replace outliers with the average value of their nearest non-outlier values [34]. Outliers were identified based on the condition that sEMG values fell outside the range from mean($v$) - 3×std($v$) to mean($v$) + 3×std($v$), where $v$ refers to a vector of 4-s sEMG signals, mean($v$) and std($v$) refer to the mean value and standard deviation of $v$. Lastly, the segmented signals were concatenated in their original order to form the continuous original signals. Fig. 4 (d) and (h) depict a representative sEMG signals of the voluntary contraction, and its enlarged areas ranging from 0 to 0.25 s, respectively.

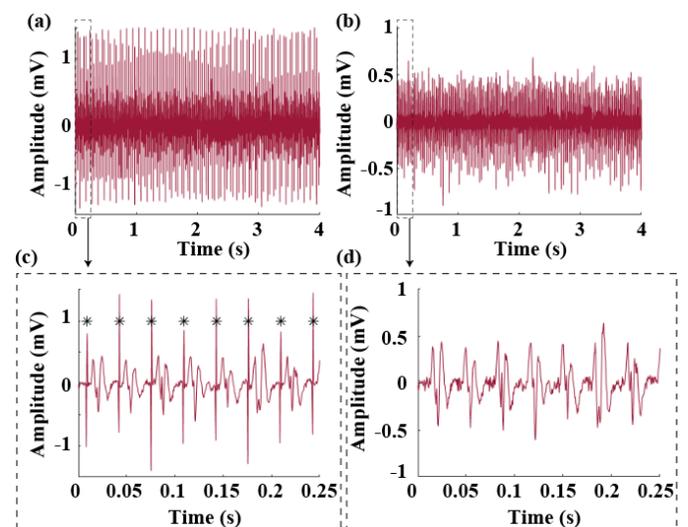

Fig. 3. LF stimulation artifact removal procedure. (a) presents 4-s unprocessed sEMG signals. (b) shows the sEMG signals from (a) after the LF stimulation artifact removal procedure. (c) and (d) are the enlarged areas of (a) and (b) ranging from 0 to 0.25 s for a clear demonstration of sEMG variations, respectively. The red asterisks in (c) indicate the identified positions of LF stimulation artifacts.

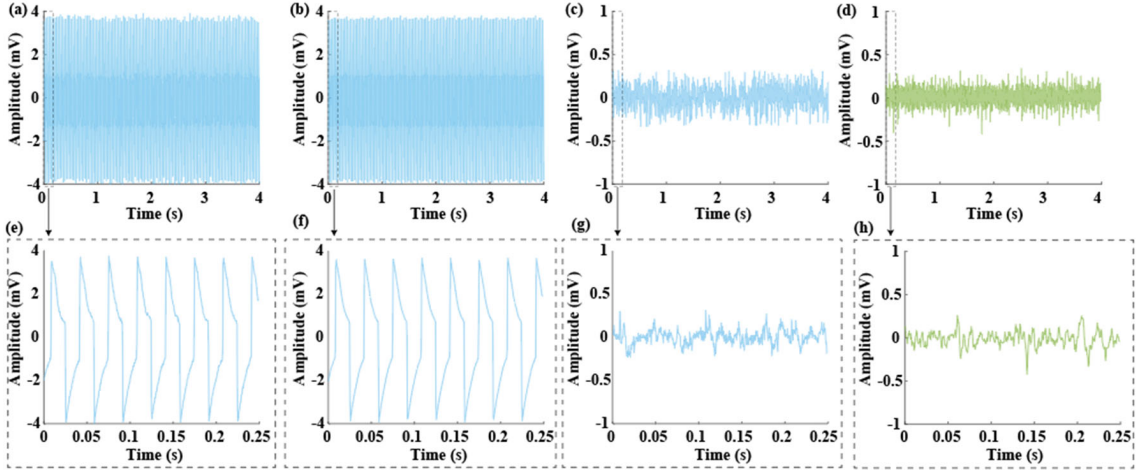

Fig. 4. HF stimulation artifact removal procedure and representative sEMG signals during voluntary contraction. We processed the HF stimulation artifact for each 4s-duration sEMG signal. (a) illustrates a 4-s unprocessed sEMG signals. (b) describes the concatenated HF artifact. (c) shows the sEMG signal after HF stimulation artifact removal. (d) presents 4-s sEMG signals during voluntary contraction, recorded under the same time interval and target force conditions as those used for both HF and LF stimulation signals. (e), (f), (g) and (h) are the zoom-in windows of (a), (b), (c) and (d) ranging from 0 to 0.25 s for a clear demonstration of sEMG variations, respectively.

After the stimulation artifact removal procedures, we compared root mean square (RMS) values of different experimental conditions (voluntary activation, HF stimulation, and LF stimulation), because the RMS value is a preferred feature to measure the sEMG amplitude, effectively reflecting the average electrical activity of the muscle. Specifically, we calculated the average RMS values during the same time periods as the force analysis. To ensure a fair comparison across different durations, for each period, we segmented the sEMG signal into consecutive 1-s segments and calculated the average RMS values for these segments.

*D. Statistical Analysis*

We conducted the statistical analysis using the parametric tests (Repeated Measures Analysis of Variance (RM ANOVA) and t-tests) when the compared groups met the primary criteria: 1) Gaussian distribution ($p > 0.05$, indicated by Shapiro-Wilk test), and 2) sphericity assumption ($p > 0.05$, indicated by Shapiro-Wilk test), as indicated by the Mauchly's test [35], [36]. If these criteria were not met, we employed non-parametric tests (the Friedman test and Wilcoxon signed-rank test) for statistical analysis [37]. For post-hoc comparisons, we applied the Holm-Bonferroni correction to address multiple-comparison errors. In this study, only the adjusted p-values were reported, and we performed the statistical analyses in MATLAB with an $\alpha$ value of 0.05 [38].

## III. RESULTS

To verify that the subjects had fully recovered from muscle fatigue before each trial, we analyzed the MVC forces before each trial. As shown in Fig. 5(a), no significant differences were detected among the three experimental conditions (Vol, HF, and LF) prior to each trial of different force levels (all $p > 0.05$), indicating muscle strength from previous trials had recovered effectively. Additionally, we compared the initial finger forces (average finger forces during the initial 5-15 s) to ensure that all subjects started trials with a matched force level across experimental conditions. The analysis revealed no significant differences (Fig. 5(b), all $p > 0.05$) in initial forces.

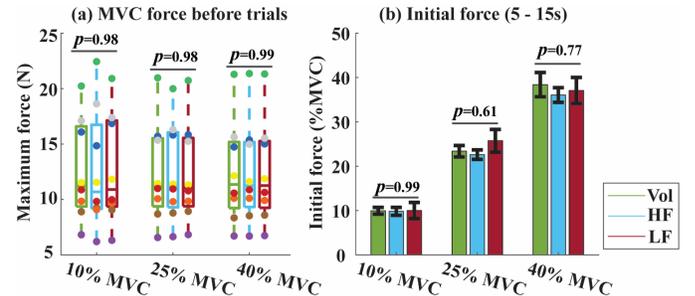

Fig. 5. (a) Muscle fatigue recovery reflected by the maximum force prior to the experimental trial and (b) Initial force comparison at the beginning of the trial. Error bars represent standard errors.

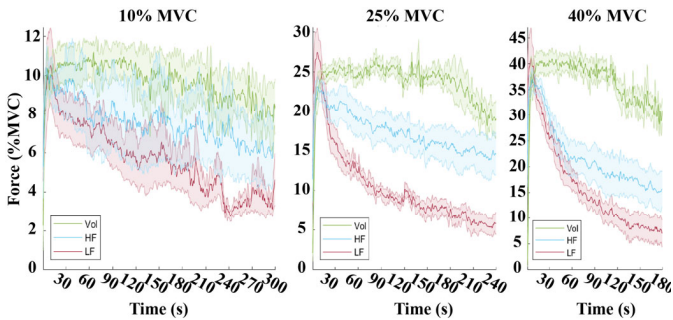

Fig. 6. Finger force profile under three force-evoked strategies for three force levels. Error bars represent standard errors. * denotes 0.01<p <0.05, ** denotes 0.001<p <0.01, *** denotes p <0.001.

TABLE II
RESIDUAL FORCES DURING THE LAST TIME PERIODS (MEAN ± STANDARD ERROR %MVC)

| Force level | Vol | HF | LF |
|---|---|---|---|
| 10 % MVC | 8.73±1.17 | 6.54±1.74 | 3.71±0.59 |
| 25 % MVC | 21.99±1.49 | 15.00±2.37 | 6.52±1.09 |
| 40 % MVC | 32.47±2.15 | 16.85±3.46 | 8.47±2.46 |

### A. Finger Forces

Fig. 6 shows the force profile (averaged across subjects) over time for the voluntary contraction, LF stimulation, and HF stimulation at the three force levels. It was evident that the HF stimulation had a distinct advantage in maintaining force output compared with the LF stimulation. Across all three force levels, although the force gradually declined over time, the force evoked by the HF stimulation remained higher than that from the LF stimulation, and the rate of decrease was slower. This illustrated the effectiveness of the HF stimulation in sustaining force output over extended duration. Additionally, the finger force from the voluntary contraction tended to be stable initially and then gradually declined over time for each force level. Compared with the LF stimulation, the force evoked by the HF stimulation was closer to the voluntary contraction force in terms of final force amplitude and force decline rate, indicating its potential for enhanced performance and efficacy in muscle fatigue reduction.

To make quantitative comparisons, we presented the average forces over different time periods (Fig. 7). For each force level, the force of the HF stimulation was consistently higher than that of the LF stimulation, and closer to the voluntary contraction force. For the 10 % MVC force level, no significant difference was detected for forces evoked by the three stimulation strategies over different time periods (all $p > 0.05$). For the 25 % MVC and 40 % MVC force levels, the voluntary contraction force was significantly higher than that of the HF stimulation in most cases (all $p < 0.05$), and the HF force in turn significantly surpassed the LF force across most time periods (all $p < 0.05$).

As time progresses, a gradual decline in muscle strength can be observed, indicative of muscle fatigue. Notably, the residual forces during the last time periods offer critical insights into muscle fatigue after prolonged muscle activation. The residual forces are presented in Table II. For both the 25 % MVC and 40 % MVC force levels, the residual voluntary contraction force was the highest among the three experimental conditions (all $p < 0.05$). The residual force of the HF stimulation was significantly greater than that of the LF stimulation (25 % MVC: $p = 0.010$; 40 % MVC: $p = 0.036$), suggesting that the HF stimulation was more effective in maintaining muscle force output over extended periods compared with the LF stimulation.

### B. sEMG Activity

The RMS value of the sEMG signal is an important indicator of muscle activity. Changes in RMS values can reflect the electrophysiological changes in muscle activity. Fig. 8 depicts the RMS variation under the three experimental conditions for each force level. The RMS values under the LF stimulation were the highest among the three experimental conditions over different time periods. The RMS values under the HF stimulation were significantly lower than those under the LF stimulation in most cases (all $p < 0.05$). Additionally, no significant difference was detected between the RMS values under the voluntary contraction and the HF stimulation (all $p > 0.05$). This suggested that the HF stimulation and voluntary contractions show similar electrophysiological profiles, which were different from the LF stimulation.

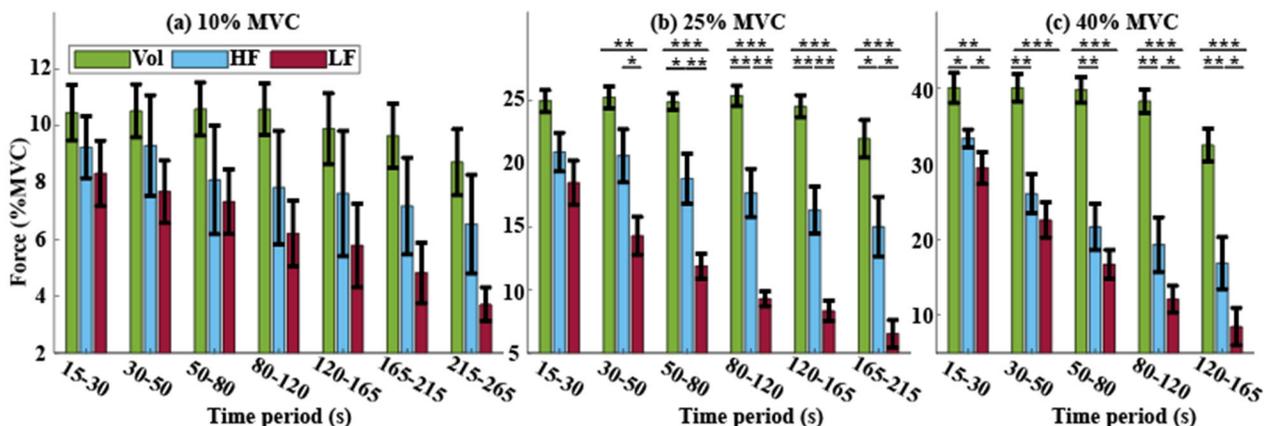

Fig. 7. Finger force performance during different time periods with the three experimental conditions for (a) 10 % MVC force level, (b) 25 % MVC force level, and (c) 40 % MVC force level. Error bars represent standard errors. * denotes 0.01<p <0.05, ** denotes 0.001<p <0.01, *** denotes p <0.001.

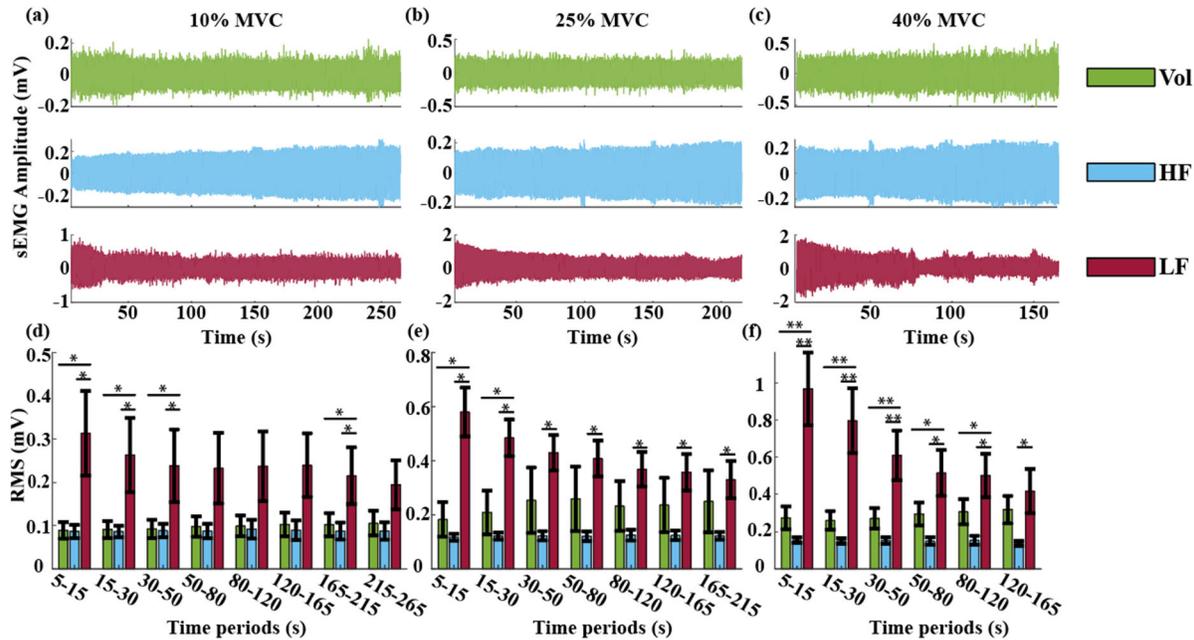

Fig. 8. Representative sEMG signals and RMS values under the voluntary contraction, HF stimulation, and LF stimulation. (a), (b), and (c) illustrate the representative sEMG signals recorded at (a) 10 % MVC condition, (b) 25 % MVC condition, and (c) 40 % MVC condition, respectively. Similarly, (c), (d), and (e) depict the corresponding sEMG RMS values under (a) 10 % MVC, (b) 25 % MVC, and (c) 40 % MVC conditions, respectively. Error bars represent standard errors. * denotes 0.01<p <0.05, ** denotes 0.001<p <0.01.

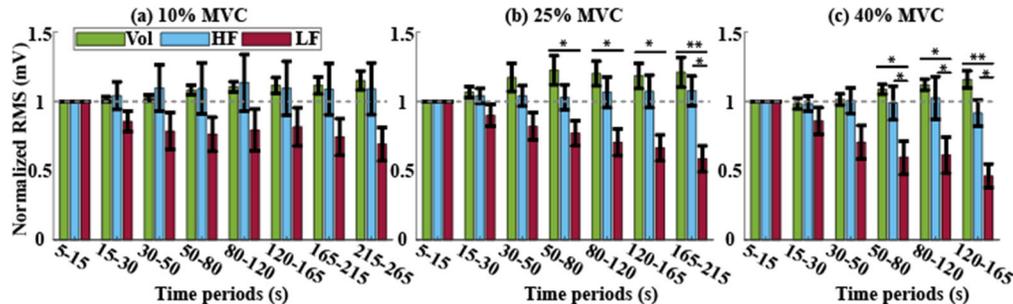

Fig. 9. Normalized RMS under the voluntary contraction, HF stimulation, and LF stimulation for (a) 10 % MVC force level, (b) 25 % MVC force level, and (c) 40 % MVC force level. The horizontal dashed lines at 1 were presented to offer an intuitive comparison with the initial states. Error bars represent standard errors. * denotes 0.01<p <0.05, ** denotes 0.001<p <0.01.

As shown in Fig. 9, we normalized all RMS values by the RMS values of the initial muscle activation (time period: 5-15s), allowing for a direct comparison of muscle activation across different time periods and conditions. The normalized RMS values under the LF stimulation showed a trend of gradual decrease over time. In comparison, the RMS values under the HF stimulation tended to be stable, closely resembling those under the voluntary contraction. Statistical analyses revealed no significant differences in normalized RMS values between the HF stimulation and voluntary contraction (all $p > 0.05$). At 25 % MVC, the normalized RMS values under the HF stimulation were significantly higher than those under the LF stimulation during the final time period ($p =0.017$). Similarly, at 40 % MVC, significant differences were observed in the time periods of 50-80 s, 80-120 s, and 120-165 s (all $p < 0.05$).

## IV. Discussion

In this study, we evaluated the efficacy of muscle fatigue reduction of a continuous, subthreshold kHz stimulation approach targeting the peripheral nerves. Compared with the conventional LF stimulation, our results revealed that the HF stimulation can effectively reduce muscle fatigue after sustained stimulation that evoked three force levels. The force decay and muscle activation pattern were closer to that under the voluntary contraction. The effective muscle fatigue reduction by the HF stimulation can allow more prolonged motor output in daily tasks that require sustained muscle activation, such as holding or transporting objects. The outcomes can provide confidence during the implementation of this FES technique in neural assistance and rehabilitation therapies.

The electrically evoked forces decreased initially and then gradually stabilized under both stimulation strategies. This trend was especially evident at high force levels. With matched initial forces, the force evoked by the HF stimulation declined slower than that of the LF stimulation over time, and the residual forces at the end of the stimulation trial was also higher than that of the LF stimulation. These results suggest that the HF stimulation can effectively reduce muscle fatigue to maintain the force output compared with the LF stimulation. The slower force decay under the HF stimulation could be because the muscle fibers from various MUs were less synchronous in this stimulation. Note that we did not perform a regression using an exponential decay function on the force profile, because a large number of the trials followed a linear trend instead of an exponential decay trend, especially in the HF and voluntary conditions. As shown in Fig. 2(b), an individual LF pulse can potentially activate a set of MUs with different magnitudes of subthreshold depolarizations. Correspondingly, the small delay between activations of different MUs leads to a highly synchronous contraction, and different MUs would discharge at the fixed rate governed by the stimulation rate. In contrast, the dispersed activation of MUs under the HF stimulation (Fig. 2(c)) is characterized by increased delays between the recruitments of axons of different diameters. Specifically, a single HF pulse can only evoke a subthreshold depolarization. Various axons need multiple depolarization current pulses to accumulate before reaching the threshold for activation, resulting in asynchronous activation. The summation of multiple sub-threshold depolarizations would allow different MUs to fire at different rates based on their intrinsic properties as shown in the earlier study [7].: The larger diameter axons in the nerve bundle are likely afferent fibers. As shown in Fig. 3(c) and 3(d), a zoom-in view of the muscle compound action potentials demonstrated substantial H-reflex activation in the LF stimulation condition. We expect that the HF stimulation protocol would recruit substantial afferent fibers and that the muscle activation is driven substantially by the reflex pathways. This would lead to lower threshold MUs being recruited earlier and firing at higher rates than the higher threshold MUs. If a reverse recruitment order were to occur, one would expect that the low stimulation intensity (10 % MVC) would have induced a faster rate of force decay than the high intensity stimulation (40 % MVC). Nonetheless, our current EMG recordings showed highly asynchronous activities, which made it difficult to verify the reflex activation. Future studies, potentially through computational models, are needed to evaluate the relative contribution of direct motor axon vs. reflex recruitment patterns.

The dispersed activation was also evident from the low amplitude and extended peak duration of the compound action potentials in the HF stimulation trials, suggesting a low degree of alignment among asynchronous action potentials [27]. To intuitively present such features, we included a representative spatial RMS distribution of the sEMG activity under 40 % MVC force level condition over time (Fig. 10). It was evident that the RMS amplitude magnitude under the HF stimulation was less than that under the LF stimulation, as shown in the different scaling of the color bars over spatial RMS distributions. Additionally, peak RMS values within the activation area tended to be stable under the HF stimulation. However, the peak RMS values under the LF stimulation gradually decreased, which was also evident in Fig. 8. The higher RMS observed for LF stimulation compared with HF stimulation can be explained potentially by the following factors. First, the stimulation amplitude in the LF condition was significantly larger than that in the HF condition ($p < 0.001$), as shown in Table I. The difference in stimulation amplitude combined with different stimulation patterns can potentially lead to different sEMG amplitude. Second, the synchronous activation of motor units under LF stimulation resulted in effective superposition of motor unit potentials, further elevating the sEMG amplitude. These factors collectively could contribute to the higher RMS values for LF stimulation. Furthermore, similar to the voluntary contraction, the spatial RMS distribution of the HF stimulation tended to be more dispersed than that of the LF stimulation. Namely, the same force output was produced by a larger set of muscle fibers in the HF stimulation compared with the LF stimulation, effectively reducing muscle fatigue.

It is well-known that force output decreases after fatigue onset [13]. However, the finger force under the voluntary contraction tended to be stable initially and then gradually declined as the time progressed for each force level, primarily because subjects voluntarily exerted more effort when fatigue occurred. This additional effort helped them maintain the initial level of force despite the increasing fatigue. Essentially, as fatigue sets in during the tasks, the capacity of muscles in generating forces declines. To compensate for this reduction and to sustain performance, the increase in effort can be reflected in enhanced muscle activation as depicted in Fig. 9 and Fig. 10(a), due to recruitment of larger MUs. However, these larger MUs tend to be more fatigable. Therefore, despite an increase in the intensity of muscle activation, there is an eventual decline in voluntary force output. Such a compensation effect in the voluntary contraction leads to a mismatch with the stimulation trials where the stimulus input was maintained at a fixed level in a trial. In future studies, we plan to implement a closed-loop control of the stimulus input based on the force output, and the stimulation intensity would increase to compensation for a force decline. We expect that the HF stimulation could lead to more sustained force output and be more similar to the voluntary activations.

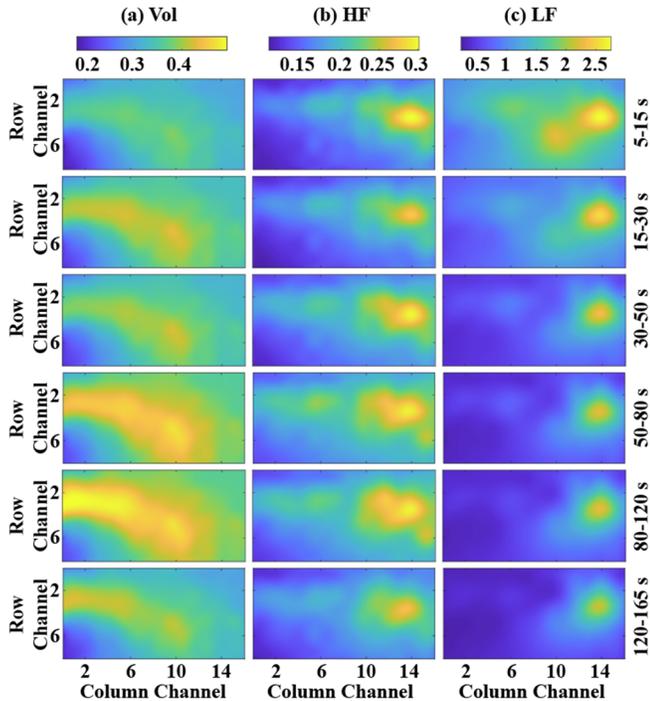

Fig. 10. Spatial RMS distribution of sEMG activity at 40 % MVC for (a) voluntary contraction, (b) HF stimulation and (c) LF stimulation over time. Each heatmap was enhanced using cubic spline interpolation to increase the point density by a factor of 10. The color scales at the top of the heatmaps represent the intensity of muscle activity. The top, bottom, left, and right of each heatmap correspond to the medial, lateral, proximal, and distal directions, respectively.

In practical FES-based applications, dynamic adjustments to the stimulation intensity are required to maintain consistent force output due to fatigue, electrode shifts, and neuromuscular adaptations. One potential approach could involve integrating closed-loop control mechanisms based on real-time force feedback [15]. Such a system could autonomously adjust the stimulation parameters to maintain the desired force levels, thus minimizing the need for manual intervention. Our findings revealed the efficacy of HF stimulation in maintaining force output and reducing muscle fatigue across different force levels, providing a foundation for future advancements in real-time adaptive control strategies. Future research could explore the feasibility of implementing closed-loop control mechanisms to enhance the robustness and practicality of HF stimulation-based force generation systems.

Our findings also have potential applicability in high-dosage muscle re-training and re-education for individuals with neurological disorders. The subthreshold kHz stimulation method can reduce muscle fatigue while eliciting MU activation patterns more like voluntary contractions than traditional LF stimulation techniques. By sustaining force output for longer durations, HF stimulation could improve the effectiveness of rehabilitation protocols aimed at restoring functional movement. This approach also has the potential to enhance neural plasticity and facilitate more natural neuromuscular adaptations, ultimately allowing extended, intensive training sessions in clinical or home-based settings.

Although the developed HF stimulation can effectively reduce muscle fatigue across a wide range of force levels, the current study has several limitations. Initially, we used a fixed pulse width of 80 μs and a fixed pulse interval of 20 μs for the HF stimulation, based on previous studies [7], [27] and the resolution constraint of our stimulator. While these studies provided valuable guidance for selecting the 10 kHz parameter, future studies can investigate the effects of other pulse width and pulse interval parameters on muscle fatigue reductions. Additionally, the stimulation was not user-specific. A detailed physical geometry model (e.g., from finite element modeling of tissue) might replicate axon firing patterns with more accurate stimulation amplitudes, we plan to develop user-specific geometric models in upcoming studies. Moreover, the current study primarily focused on time-domain features of sEMG signals. In future studies, we will conduct frequency domain analysis to further explore the fatigue effect on EMG signals during voluntary effort and our HF stimulation.

## V. Conclusion

Our study investigated the effectiveness of HF stimulation for reducing muscle fatigue across different force levels. The HF stimulation of different amplitudes was delivered using continuous, charge-balanced, subthreshold pulses at kHz frequency. The results demonstrate that HF stimulation effectively reduces muscle fatigue compared to conventional LF stimulation. Furthermore, our findings show that HF stimulation not only evokes a more dispersed and natural muscle activation pattern, closely resembling voluntary contractions. These outcomes hold promise for enhancing the practical application of FES in clinical populations, especially for individuals with neurological disorders who rely on these systems for daily muscle activation and rehabilitation.